\def\ba{ \begin{eqnarray}}
\def\ea{ \end{eqnarray}}

\def\be{ \begin{equation}}
\def\ee{ \end{equation}}

\documentstyle[aps,preprint,12pt]{revtex}

\title{ Is Quantum Mechanics Non-Local?}

\author{ W. Unruh}
\address{ Program in Cosmology and Gravity of CIAR\\Dept 
Physics and Astronomy\\University 
of B.C.\\ Vancouver, Canada V6T 1Z1}
\begin{document}
\maketitle
\tightenlines

\begin{abstract}
Stapp has recently argued from a version of the Hardy type experiments
that quantum mechanics must be non-local, independent of any additional
assumptions like realism or hidden variables. I argue that either his
conclusions do   not follow from his assumptions, or that his
assumptions are not true of quantum mechanics and can be interpreted as
assigning an unwarranted level of reality to the value of certain
quantum attributes.
\end{abstract}

In a recent paper\cite{1} H. Stapp has argued that quantum mechanics is 
non-local. 
He bases this conclusion on a weak definition of locality for a physical 
theory and an examination of the predictions of quantum mechanics for a 
specific two particle Hardy-type\cite{2} \cite{3} experiment. It was already 
known since 
the 50's from the theorem of J.S. Bell\cite{4} that quantum mechanics violated 
the correlations expected in any theory which is both local and is "hidden  
variable" theory, in which the statistics arises out of the lack of knowledge 
of the values of local hidden variables which determine the outcome of 
the experiment itself. However, Stapp's argument claims to strengthen
that conclusion by arguing that no local theory cannot give the same
predictions as quantum mechanics. Ie, he reaches the conclusion that
quantum mechanics is non-local.

In his paper, Stapp claims that his proof that Quantum mechanics is non-local 
applies only if one assumes that locality holds in all frames of a special 
relativistic system. However I will take this as a given, that locality 
is to be taken to apply to any two systems when causally separated from 
each other. The temporal order of experiments which are space-like separated  
is assumed to be irrelevant. 

To begin, let me first present the argument in my own way. Stapp  
phrased his argument in the form of a logical calculus, which I will examine 
later.
He considers a very general situation in which measurements are made at
two locations, L (for left) and R (for right). At each location two
possible measurements are made which he designates by $L1$ and $L2$ for
the left and $R1$ and $R2$ for the right. In this first section I will
specify his generic situation by having the measurements be made on the
spins of two particles, one particle labeled L on the left and the other
,R, on the right.
These two particles are 
placed into a specific correlated state. For purposes of illustration, 
I will assume that these are two spin-1/2 particles, and that in the spin-z 
basis for each particle, the state is given by
\be
|\Psi>= N\left(\cos(\theta)|+>|+> +\sin(\theta)|+>|->
+{1+\sin^2(\theta)\over \cos(\theta)}|->|+>-\sin(\theta) |->|->\right) 
\ee
where the state $ |\alpha>|\beta>$ refers to the left particle being 
in the $\alpha$ eigenstate of its spin-z operator and the right particle 
is in the  $\beta$ eigenstate of its spin-z operator.
 $\theta $ is a fixed parameter, and $N$ is a normalisation given by
$$N=\cos(\theta)/\sqrt{2 (1+\sin^2(\theta))}.$$

Writing this state in the form
\be
|\Psi>= N\left(|+>(\cos(\theta)|+>+\sin(\theta)|->) + |->({1+\sin^2(\theta)\over 
\cos(\theta)}|+>-\sin(\theta)|->)\right) 
\ee
we see that that the $|+>$ eigenstate of the left particle is perfectly 
 correlated with the eigenstate $(\cos(\theta)|+>+\sin(\theta)|->)/\sqrt{2}$ 
for the right
particle,  which is the + eigenstate of the operator $\sigma_\theta= \cos(2\theta)\sigma_z+\sin(2\theta)\sigma_x$  
for the right particle. I.e., if $\sigma_z$ is measured on the left
 particle and is found to have value +, and $\sigma_\theta$ is measured 
on the right, then it will always be found to have value +.

 Writing the state as
\ba
|\Psi>= N&&\left(  (|+>+|->)(\cos(\theta)|+>+\sin(\theta)|->)\right. \\ 
&&\left.+
 2 \tan(\theta)|->(\sin(\theta)|+>-\cos(\theta)|->) \nonumber
\right) 
\ea
we see that that the $(\cos(\theta)|+>+\sin(\theta)|->)/\sqrt{2}$ eigenstate 
of the 
right particle's $\sigma_{\theta}$ operator is perfectly  correlated with 
the eigenstate $(|+>+|->)/\sqrt{2}$ of $\sigma_x$ of the the left particle.
 I.e, if $\sigma_\theta$ is measured on the right and found to have value 
+,
 then if $\sigma_x$ is measured on the left, it is always found to have 
value +.

 Finally, 
we can write the state as  as 
\be
|\Psi>=N\left( (|+>+|->) {1\over \cos(\theta)}|+>
 +(|+>-|->) ({-\sin^2(\theta)\over 
\cos(\theta)}|+> +\sin(\theta)|->) \right) 
\ee
and we see that the state $(|+>+|->)/\sqrt{2}$ of the left particle is  
perfectly   correlated 
with the state $|+>$, the plus eigenstate of the $\sigma_z$ operator, of 
the right particle.

We thus can construct a chain of perfect correlations. If we have measured 
$\sigma_z$ of the left 
particle and found it to be $+$, then we can predict with certainty that 
the 
if we had measured $\sigma_\theta $ 
of the right, the answer would have been +. If we measured $\sigma_\theta$ 
of the right and found a value of +, then we would know with certainty 
that if we had measured $\sigma_x$ of the left we would have gotten a 
value of 
$+$. Finally if we had measured $\sigma_x$ of the left particle 
and gotten +, we would know with certainly that we would have gotten the 
+ eigenstate if we had measured $\sigma_z$ for the right particle. 

In essence, the Stapp argument is that we can use the logical chain 
\be
\sigma_{Lz}=+ \Rightarrow \sigma_{R\theta}=+\Rightarrow\sigma_{Lx}
=+\Rightarrow \sigma_{Rz}=+\label{inference}
\ee
 to infer that $\sigma_{Lz}=+\Rightarrow \sigma_{Rz}=+$ with certainty. 
However a direct calculation shows that if $\sigma_{Lz}=+$, one has only 
a probability of $cos^2(\theta)$ that if one also measures $\sigma_{Rz}$ 
that one will get a value of $+$.  If one chooses $\theta$ very near $\pi/2$, 
one can make this probability as small as one desires. Ie, the inferred 
value of $\sigma_{Rz}$, given that $\sigma_{Lz}$ is measured to have value 
+, and given the chain of inferences via the perfect correlation, is with
 high probability exactly the opposite of what quantum mechanics 
would predict would be obtained in a measurement. Stapp essentially argues 
that the truth of the  chain of inferred values can be justified by an 
appeal only to to locality. Thus, if correct,
  it is  relativistic locality alone, with no further assumptions,
which allows one to carry out the line of inferences. Since quantum
mechanics violates the conclusion of that inferential chain, his
conclusion is that quantum mechanics must be non-local.

Now, if one believed that attributes of a system had values which the measurements 
simply revealed, and if measurements made in one causally disconnected 
region 
cannot influence those values of attributes associated with some other 
causally
 disconnected region, then the truth of that inferential chain would be 
immediately obvious. 
This is essentially a strong form of hidden variables theory (where those 
values
 can be derived from the values which some set of hidden variables have 
in any 
particular realisation of the experiment), and this 
analysis shows that quantum mechanics is in disagreement with such a 
hidden variables theory. However, a weaker form of hidden variables theory 
is that such values are context sensitive. Ie, that the value of an attribute 
may depend on the types of experiments which are actually carried out on 
the system. This is where the assumption of locality comes in as a restriction 
on the types of context sensitivity of the values that attributes can have. 
In particular, locality is usually used to argue that value that a variable 
attains must be independent of the choice of experiment carried out in 
a causally disconnected region (although correlations clearly mean that 
the value need not be independent of the values obtained for measurements 
in disconnected regions). Thus the context for the value assigned to an 
attribute can depend on the experiments carried out in the causally connected 
region surrounding the experiment, but not on the experiment carried out 
in the causally disconnected region.

However within quantum mechanics, attributes do not have values unless 
those attributes are actually measured. Thus, if $\sigma_z$ of particle 
L has 
value +, it is inappropriate within quantum mechanics to argue that $\sigma_{Lx}$ 
must have some 
value. It was not measured and thus one cannot talk about the value that 
it has.
Talking about the the values of non-measured attributes is termed ``counter-factual"
 reasoning. Clearly the inference chain in eqn.\ref{inference} relies on counter-factual 
reasoning 
since within the context of quantum mechanics, only one of a non-commuting 
set of
 operators can be measured at any one time. One attitude toward any such 
chain of
 argument would be to disallow all counter-factuals. However, as Stapp 
argues,
 such a procedure would be to disallow reasoning which physicists often 
engage 
in, even in quantum situations. His example is classical, where such counter-factual
 reasoning is unexceptional, but quantum examples could also be found. 
However, in 
the quantum case one must be extremely careful in carrying out such arguments, 
and 
must ensure that one is not assuming a form of quantum realism-- that quantum 
attributes
 have values even if they have not been measured-- together with the 
counter-factual discussion.

 I will assume that such counter-factual statements may legitimately be
made in certain circumstances. Given that one has established a
correlation of system A (which since I am making a generic argument I
will use instead of say L of the above specific example) with system B
(instead of the specific R from above), then one can make measurements
on system B, and on the basis of
 the known correlations, make inferences about system A, even if system
 A has not been directly measured. After all, if
such
 reasoning were disallowed, the whole of the von Neuman argument about
 measurement would be invalid. In von Neuman's discussion, it is
 precisely the use of correlations of measuring apparatuses with
systems that allows us
 to deduce properties of the system from measurements made on the
 measuring apparatus, even though no direct ``measurement" has been
made on the system.

However, great care is required in such counter-factual statements that
one
 does not import into the statements a notion of reality. In
 particular, the truth of the statement made about system A which
relies on measurement made
 on system B and on the correlations which have been established
 between A and B in the state of the joint system is entirely dependent
on the truth of
 the actual measurement which has been made on system B. To divorce
 them is
 to effectively claim that the statement made about A can have a value
in
 and of itself, and independent of measurements which have been made on
A.
 This notion is equivalent to asserting the reality of the statement
 about A independent of measurements, a position contradicted by
 quantum mechanics.

Thus in the above system, measuring $\sigma_\theta$ on particle R and
finding value + can lead one to assigning a value of + to $\sigma_x$ of
particle L, even if that attribute  was not directly measured, due to
the correlation between the two particles.  However, that value for
$\sigma_{Lx}$ is entirely dependent on the fact that $\sigma_\theta $
was actually measured and found to have a certain value on
particle R. In particular, causality cannot be used to argue that the
inferred (as opposed to measured)
 value of $\sigma_{Lx}$ must be independent of what was measured at
particle R. Although the two measurements may be causally disconnected,
they are not logically disconnected. The value can be assigned to
$\sigma_{Lx}$ is logically tied to the actual measurement of
$\sigma_{R\theta}$ and its value.

Without the extension of the concept of locality to such inferred
values, I will argue below that the chain of reasoning used by Stapp to
establish eqn.\ref{inference} is broken. In all cases at most one
attribute (either $\sigma_{Lz}$ or $\sigma_{Lx}$ for the left particle
or $\sigma_{Rz}$ or $\sigma_{R\theta}$ for the right particle) is
measured at each of the particles. That measurement may be used to
infer some counter-factual value at the other particle, but in each
case that chain of inference cannot be extended sufficiently to obtain
the conclusion of eqn.\ref{inference} .       Since his argument breaks
down, there is no contradiction between the statement that quantum
mechanics is local--ie, that measured (as opposed to inferred) values
must be independent of the attribute measured in a causally
disconnected region-- and that the quantum outcomes obey the rules that
they do.

Let me finally examine Stapp's argument in detail. Stapp postulates
three requirements that he claims a local theory should satisfy. He
gives labels to particles, variable and outcomes, and uses the notation
of a logic calculus.  He refers to two measurement situations, L and R
which are space-like separated (ie, causally disconnected). These would
correspond to my  particles each with spins on the left and right
mentioned above.
 Thus in order to make contact with his notation, I will assume that
the measurement of $\sigma_z$ at L is designated by his specific
measurement situation on the left, L1, while measuring $\sigma_x$ at L
is designated by his L2. Measuring
$\sigma_z$ on particle R is R1 and $\sigma_\theta$  on the particle R
is R2.  His outcomes $a$ and $b$  are mapped onto my obtaining the
values $-$ and $+$ on L1 or, more specifically, on obtaining the values
$-$ and $+$ on a measurement of $\sigma_z$ on the particle at L.
Similarly, $c$ and $d$ refer to obtaining the value of $+$ and $-$ on
$L2$ ($\sigma_{Lx}$). $e$ and $f$ refer to obtaining values $-$ and $+$
on $R1$, ($\sigma_{Rz}$) and $g$ and $h$ to values $+$ and $-$ on R2
($\sigma_{R\theta}$). By assumption, L and R are measured in causally
disconnected regions. Thus in all cases one can assume that L (R) is
measured after any measurement is made of R (L) and the free choice of
which measurement is actually made on L (R) should have no impact
on the outcomes of measurements made on R (L).

His locality conditions, in his modal logic calculus are:
\be
LOC1: Ru\wedge Lv\wedge i\Rightarrow Ru'\Box\rightarrow Lv\wedge i
\ee
or in words, a change in the choice of which experiment the experimenter 
at R 
will   carry out does not affect the outcome of the experiment at L where 
Lv is measured.
 $u$ refers to one of 1 or 2, and u' refers to the other of the two.
$v$ refers to 1 or 2, and $i$ refers to one of the possible outcomes of
the measurement $Lv$. $\wedge$ means ``and" and the symbol
$Ru'\Box\rightarrow$ means "if Ru is replaced by Ru' in the previous
expression". This definition of locality is unexceptional if we limit
ourselves to $Lv\wedge i$ actually measured. I discuss this further
below when I go through Stapp's argument step by step.

\ba
LOC2:&& {\bf If} \left( Lv\Rightarrow \left[ (Ru\wedge j)\Rightarrow (Ru'\Box\rightarrow 
Ru'\wedge j'\right]\right)\\
&&{\bf Then} \left( Lv'\Rightarrow \left[ (Ru\wedge j)\Rightarrow (Ru'\Box\rightarrow 
Ru'\wedge j'\right]\right)
\ea

This says that {\bf If}, under the condition that Lv is measured (at a 
late time) at
 L, one can deduce that some condition prevails relating measurements only 
on 
the right hand side, {\bf Then} that relation on the right must also be 
true if the 
experimentalist on the left side had decided to make some other
 measurement (at that later time). I.e.,  the truth of a statement pertaining 
exclusively to possible events in region R cannot depend on which free 
choice 
is to be made by the experimenter in region L.

 I have inserted the "at a 
late time" to emphasise that these two measurements are causally unrelated
 and thus one can assume that L is measured at a time later than R.   Note
 that in LOC1, it is R that is taken to be the later, while in LOC2 it 
is
 L. Under the notion of relativistic causality, this is a completely
 legitimate procedure. Either of two causally separated events can be 
taken to be the earlier in relativity.
In his paper, Stapp questions the legitimacy of this procedure. I do not,
 and see no possible reason for doing so. In my opinion, if quantum 
mechanics is to be a local theory it should be local under the full 
relativistic definition.  
 
LOC2 can be rewritten as
\ba
LOC2:&& (Lv\wedge Ru\wedge j\Rightarrow Ru'\Box\rightarrow Lv\wedge Ru'\wedge 
z)
\\
&& \Rightarrow Lv'\Box\rightarrow 
Ru\wedge j (Lv'\wedge Ru\wedge j\Rightarrow Ru'\Box\rightarrow Lv'\wedge 
Ru'\wedge j)\nonumber
\ea
To me this more clearly states the content of the above principle and
 says that if from the fact that Lv is measured and Ru has 
the value j you can infer that if Ru' had been measured instead it would 
have had value j' then that inference must by independent of whether you 
measured Lv or Lv'.  

If it were true that one could deduce solely from the fact that a measurement 
had been made
at L that some relation on the right hand side must hold, then I would 
agree that 
this requirement would be reasonable. However, if the truth of the relation 
on the right
hand side depended not only on which measurement had been made on the left, 
but {\bf also} 
on the actual value obtained on the left, then no such locality relation 
would obtain.  If it
is the value obtained on the left, even if that value is obtained at a 
later time, which allows
one to deduce the relation on the right, then that relation on the right 
cannot be independent of
what it is that is measured on the left, but rather is tied to that measured value. 
To assume otherwise, to assume that the  relations between possible
 measurements on the right are
independent of the  values of the outcomes on the left which were used
to derive those relations  is, in my opinion,
simply another form of realism.  
 
Let us now look in detail at Stapp's argument. Again I will use his
notation.  
\be
(1) LOC1: (R2\wedge L2\wedge c)\Rightarrow(R1\Box\rightarrow(R2\wedge L2\wedge 
c))
\ee
On face value this is just the unexceptional statement that if L2 is
measured to have value c, then the truth of having obtained that value
is independent of what is (or will be) measured at R. Ie, if you had
thought that R2 was measured at R and you knew that the outcome of the
measurement of L2 had been c, then you would never find that outcome to
be surprising if you were
 told that it had actually been R1 which had been measured at R. Which
 measurement was done at R has no effect on the outcome of a
 measurement made at L.

However, this meaning of LOC1, although certainly true in quantum
mechanics, is insufficient to derive his conclusion, since it demands
that L1 had actually been measured and  had the given outcome. It ties
the meaning of this locality assumption to the actuality of the
measurement and its outcome. This does not allow counter-factual
replacement of for example L, since the truth of this statement is,
under this interpretation tied to the truth of the measurement of the
attribute  L1 actually having been carried out on L, and on having
obtained that specific outcome.

Another possible meaning, is that if we carried out another experiment
in which we kept all of the conditions on the experiment the same
(except the outcomes of the measurements and which measurement we made
at R) then we would obtain the same value for L2.  This interpretation
is clearly wrong about quantum mechanics, since the outcome of any
measurement is a  random process under-determined by the conditions of
the experiment, unless that outcome is a certainty.

Finally, and this is what I suspect Stapp means,
 this statement could be taken to mean that if one is somehow able to
infer that L2 has value c, then it remains true that L2 has value c
under replacement of R1 with R2 even if the outcomes of the measurement
made of R1 was crucial in drawing the inference that L2 has value c.
This interpretation of LOC1 is, I would argue a form of realism, in
that it claims that the value to be ascribed to L2 is  independent of
the evidence used to determine that value. Clearly this
is true of a hidden variables theory, in which L2 has a value
independent of any evidence used to determine that value. The evidence
simply reveals the value. However, I would however claim that under
this interpretation, LOC1 is not true of quantum mechanics. If the
value obtained in a measurement of  R2 was crucial in determining that
L2 had value c, then R2 cannot be replaced by R1, even if it is
causally disconnected or comes later in time. If the measurement of R2
was not necessary (logically) in determining the value of L2, then of
course  the acausal relation also makes the value of L2   physically
independent of R2 and it can be replaced by R1 without changing the
conclusion that L2 has value c.

Since this is key to my argument below, let me restate this. In quantum
theory, one can determine the value of an operator in many ways. One
can directly measure that value (where the term ``measure" is taken as
a primitive of the theory of quantum interpretation) or one can infer
that that value because one has dynamically correlated that system (in
this case the L particle) with some other system (a measuring
apparatus) and then measured something on that second system. (This is
the essence of the von Neuman analysis, where he showed that under
certain conditions, the second type of measurement with an apparatus
was equivalent in its predictions with the first type.)
In the example examined by Stapp, there are two possible meanings to
the term $L2\wedge c~(\sigma_{Lx}=+)$. One can either directly measure
the value of L2 ($\sigma_{Lx}$) and find it to be c (+), or one can,
because of the correlations between the two particles, regard R as a
measuring apparatus for L. Obtaining the value g (+) for
R2($\sigma_{R\theta}$) can then be used to infer the value of c (+) for
L2 ($\sigma_{LX}$). In the former case, the value of L2 obtained by the
measurement is clearly independent of which measurements are carried
out in the causally disconnected region, and LOC1 applies. If however
the value of L2 is only inferred from the value obtained from the
``measuring apparatus"  R, then that value is clearly not independent
of the measuring apparatus, R. Ie, if in LOC1 the value obtained for L2
($\sigma_{LX}$)is inferred from some measurement on R, then LOC1 is
false, no matter what the causal relation is between L and R.

\be
(2) QM: (L2\wedge R2\wedge g)\Rightarrow (L2\wedge R2\wedge c)
\ee

This is to encompass the claim that if we know that both L2
($\sigma_{LX}$) and R2 ($\sigma_{R\theta}$) were measured, and we know
that R2 had value g (+), then L2 must have had value c (+).
 It is however important to note that this statement can have two possible 
interpretations. In one the value of c for L2 was inferred from the initial 
state and the measured value of R2, while in the other case L2 was measured 
and actually found to have value c. In that case the $\Rightarrow$ must 
be interpreted as 'it is consistent with' rather than 'it is inferred that'.

\be
(3) QM: (L2\wedge R1\wedge c)\Rightarrow (L2\wedge R1\wedge f)
\ee
This is a similar statement to the previous onei for a different
implication which can be drawn from the Hardy state.

\be
(4) LOGIC: (L2\wedge R2\wedge g)\Rightarrow (R1\Box\rightarrow(L2\wedge 
R1\wedge f)
\ee

The claim is that this arises solely by logic out of the substitution of (2) and (3) into (1).
However it is now clear that the meanings ascribed to LOC1 and to the statements in the logical calculus arecrucial. To fill in the steps of this reasoning, I get 
\ba
(L2\wedge&& R2\wedge g)\Rightarrow L2\wedge c \wedge R2\wedge g)\Rightarrow(L2 \wedge c \wedge R2)
\\
&&\Rightarrow (R1\Box\rightarrow (R2\wedge L2\wedge c)\rightarrow L2\wedge c \wedge R1\Rightarrow L2\wedge c \wedge R1\wedge f\Rightarrow L2\wedge R1\wedge f
\nonumber
\ea
There are now a number of meanings which can be attached to this sequence of statements.

In the first, L2 is assumed to have value c only  as an inference drawn
from the fact that R2 had value g. L2 is not independently measured and
found to have value c. Thus the third step in this expansion does not
now follow since L2 has value c only because R2 has value g. Within
quantum mechanics, one cannot assume the truth of a statement (L2 has
value c) independent of the evidence for that statement (R2 has value
g). One cannot therefor use LOC1 to replace R2 by R1, since one no
longer has any independent evidence that L2 has value c.   Were there
some independent evidence for the value of L2 (a direct measurement, or
a correlation between L2 and some other system such that knowing the
outcome for that other system would imply the value for L2) then one
could use LOC1 to carry out the (counter-factual) replacement of R2
with R1.

The second interpretation of this statement is that  L2 has actually
been measured at L. If L2 was actually measured, than the first four
inferences follow, but o  the last inference is no longer true. The
fact that L2 was measured and had value c and that R1 was measured and
had value f does not imply that L2 was measured and that R1 was
measured and had value f independent of the value measured for L2. The
truth of R1 having value f is an inference drawn from the actually
measured value of L2, and is not itself an independently measured value
(it could not be since R1 is counterfactual). Just as in my previous
comments about LOC1, the value associated with R1 cannot be seperated
from the evidence used to ascribe that value to it, as the last
inference attempts to do (ie, it claims that counterfactual R1 has
value f even if it is not asserted that L2 has value c, but only that
L2 was measured).

Again, it is the fact that in quantum mechanics there is a difference
between the value of an attribute as inferred from the value of some
other attribute, and the value as directly measured which breaks this
logical chain. Stapp
 must use the independence of the value of an attribute from the way in
 which that value was inferred to carry through his argument. His
 logical calculus does not distinguish between the various ways a value
 for an attribute can be determined, and thus assumes that that value
 is independent of that determination, an unwarranted assumption in
 quantum theory, and at least for me, an assumption of realism for the
 value of an attribute.

It is on this expression (or rather on a slight rewriting of this expression) 
that Stapp now uses LOC2.
\be
(5)LOGIC: (L2)\Rightarrow( R2\wedge g\Rightarrow R1\Box\rightarrow R1\wedge 
f)
\ee
This is a rewriting of (4). Again however, this expression does not capture 
the essence of (1),(2), and (3) where it would only be 
the fact of L2 actually having 
been measured to have the value c which allows us to make these
inferences, and not simply that L2 has been inferred to have value c
from measurements made on R2, or that R1 had value f inferred from the
measured value of c for L2. It would be true if the proposition $(L2)$
were replaced with $(L2\wedge c)$, where this would be interpreted as
``L2 actually having been measured and found to have value c" but this
would again not allow him to use his LOC2 to draw his ultimate
conclusion.

\be
(6) LOC2: (L1)\Rightarrow(R2\wedge g\Rightarrow R1\Box\rightarrow R1\wedge 
f)
\ee
This is the statement that if one has derived some expression about R which 
is true, then that expression should be independent of what it is that 
has been measured on the left hand side (e.g., that measurement on the 
left 
hand side could have been made long after whatever measurement had been 
made on the right.) However this neglects the fact that the inference 
was made entirely based not only on the fact that L2 was measured on the 
left, but also that the value obtained for L2 was in fact the value c. If R2 
was measured, then R1 was not in fact measured. The whole basis for the 
inference that R1 had value f is that L2 was measured to have value c. 
Again, if the value of R1 were something which was real, something which had an existence 
independent of the means used to determine what it was, then this substitution 
would make sense. But in quantum mechanics, the value is not independent 
of the means. The inference of the value of R1 is entirely based on the 
truth of the statement that L2 was measured to value c. Thus it is not 
simply locality which enters into LOC2 but also some notion of the independent 
reality of the value of R1, even if it has not itself been directly measured.

The truth of statement 4 ultimately rests on the additional
 assumption that the value of an attribute (eg, R1 ($\sigma_{Rz}$) being f (+)),
  is independent of the evidence --
L2 being g or $\sigma_{Lx}=+$ --which was used to infer that value for
R1. R1, being counterfactual, cannot have actually been measured in
some independent way. It's value is only inferred from the value
obtained for L2, and as such is not independent of that value, even if
L2 was only measured much later. That assumption of independence of
value from evidence is, I would claim, the heart of realism, and is
contrary to quantum mechanics.

His analysis does raise interesting issues in our understanding of
quantum mechanics. As mentioned above the von Neuman analysis of the
measurement process does assert that one can use the correlations
between systems ( a measuring apparatus and some system one is
measuring) to make statements about the system that is being measured.
(If one sees the pointer on a properly constructed and operating volt
meter point to 10V, then one can use that to infer that the system one
is measuring the voltage of has 10V as its value.) To deny the use of
such inferences drawn from such correlations between a measuring
apparatus and the measured system would be to destroy our theory of
almost all physical measurements. However, the Stapp argument shows that
one must treat such inferential measurements carefully. Once one has used the
measuring apparatus to infer the value of some measured quantity, one
must keep in mind that that inferred value depends crucially on that
process used to carry out the measurement, and is not independent of
that measuring apparatus. Although usually one can get away with an
elimination of the measuring apparatus and regarding the inference of a
value as being equivalent to the direct measurement of that value
(measurement here being used in its primitive sense in the
interpretation of quantum mechanics), there are   situations, as
in these Hardy type experiments, in which such sloppiness about 
the meaning of measurement can lead one
into error.

\acknowledgements
I would like to thank the NSERC for support during the course of this work 
via
 research grant, and the Canadian Institute for Advanced Research for their 
support.
 I would also like to thank H. Stapp very much for numerous discussions, both in person and by email, about the 
topic
 of this paper. Although we did not come to agreement, these discussions 
both
 helped to make clear to me exactly what his argument was, and helped to 
clarify my own thinking on the topic.

\end{document}